\documentclass[12pt,twoside]{article}
\usepackage{fleqn,epsfig,espcrc1}
\usepackage{graphicx}
\usepackage{slashed}
\def\be{\begin{equation}}
\def\ee{\end{equation}}
\def\br{\begin{eqnarray}}
\def\er{\end{eqnarray}}
\def\bc{\begin{center}}
\def\ec{\end{center}}

\def\ps {\not\!p}

\def\ps {\not\!p}

\def\M {{{\cal M}}}

\def\L {{{\cal L}}}

\def\N3{{1\over (2\pi)^3}}
\def\rf#1{{(\ref{#1})}}

\def\g5{\gamma_5}

\def\gmmunnu{g^{\mu\nu}}

\def\gmunu{g_{\mu\nu}}

\def\endauthors{}
\def\authors#1\endauthors{#1}

\def\gmmunnu{g^{\mu\nu}}

\def\gmunu{g_{\mu\nu}}

\def\gmmunnu{g^{\mu\nu}}

\def\g {{\gamma}}

\def\endauthors{}
\def\authors#1\endauthors{#1}

\def\be{\begin{equation}}
\def\ee{\end{equation}}
\def\br{\begin{eqnarray}}
\def\er{\end{eqnarray}}
\def\brn{\begin{eqnarray*}}
\def\ern{\end{eqnarray*}}
\def\rf#1{{(\ref{#1})}}

\def\ps {\not\!p}

\def\lbar{\mbox{$\lambda$\kern-0,450em \vrule width0,35em height1,252ex
depth-1,21ex \kern0,051em}}
\def\dbar{\mbox{d\kern-0,347em \vrule width0,3em height1,252ex depth-1,21ex
\kern0,051em}}
\def\Dbar{\mbox{D\kern-0,735em \vrule width0,3em height0,86ex depth-0,81ex
\kern0,40em}}

\def\g {{\gamma}}

\def\L {{{\cal L}}}

\def\M {{{\cal M}}}
\def\N {{{\cal N}}}

\def\ba#1{\begin{array}{#1}}
\def\ea{\end{array}}

\def\bc{binomial coefficients }

\def\be{\begin{equation}}
\def\ee{\end{equation}}
\def\br{\begin{eqnarray}}
\def\er{\end{eqnarray}}
\def\brn{\begin{eqnarray*}}
\def\ern{\end{eqnarray*}}
\def\bit{\begin{itemize}}
\def\eit{\end{itemize}}
\def\bnu{\begin{enumerate}}
\def\enu{\end{enumerate}}

\def\={{\simeq}}
\def\rf#1{{(\ref{#1})}}

\def\nn{\nonumber }

\def\gmp {\gamma^{\mu} }

\def\gnp {\gamma^{\nu} }

\def\2q{{{\{}2{\}}_q}}
\def\3q{{{\{}3{\}}_q}}

\def\qslash{/\!\!\!{q}}
\def\pslash{/\!\!\!{{p}}}

\def\ppslash{/\!\!\!{{p'}}}

\def\ppslash{/\!\!\!{{p'}}}
\def\eeslash{/\!\!\!{{e^*}}}

\def\psimu{\psi_\mu}
\def\psinu{\psi_\nu}
\def\barpsimux{\overline{\psi}_\mu(x)}

\def\psinu{\psi_\nu}

\def\gmp {\gamma^{\mu} }

\def\gnp {\gamma^{\nu} }

\def\onehalf{{1 \over 2}}

\def\threehalf{{3 \over 2}}

\def\gmunu{g_{\mu\nu}}
\def\gmmunnu{g^{\mu\nu}}

\def\K {{\cal K}}

\begin{document}

\begin{titlepage}
\pagestyle{empty} \baselineskip=21pt
\begin{center}
{\large{\bf About consistence between $\pi N \Delta $ spin-3/2 gauge  couplings and electromagnetic gauge invariance}}

\end{center}

\authors
\centerline{D. Badagnani$^{*}$, C. Barbero$^{**}$ and A. Mariano$^{**}$
\footnote{corresponding author:mariano@fisica.unlp.edu.ar}} \vskip -.05in
\centerline{\small \it
$^{*}$ Departamento de F\'{\i}sica, Universidad Nacional de La Plata, C.
C. 67, 1900 La Plata, Argentina}
\vskip -.05in
\centerline{\small \it
$^{**}$ Instituto de F\'{\i}sica La Plata, CONICET, 1900 La Plata, Argentina}
\vskip -.05in
\endauthors

\bigskip

\centerline{ {\bf Abstract} }
\baselineskip=18pt
\noindent

\bigskip

We analyze the consistence between the recently proposed ``spin 3/2 gauge'' interaction for the $\Delta$ resonance with nucleons ($N$) and pions ($\pi$), and the fundamental electromagnetic gauge invariance
in any radiative amplitude.
Chiral symmetric $\pi$-derivative $\pi N \Delta$  couplings can be substituted through a linear transformation to get  $\Delta$-derivative ones, which have the property of  decoupling  the 1/2 field components of the $\Delta$ propagator. Nevertheless, the electromagnetic gauge invariance introduced  through  minimal substitution in all derivatives, can only be fulfilled at a given order n without destroying the spin 3/2 one by dropping n+1 order terms within  an effective field theory (EFT) framework with a defined  power counting.
In addition,  we show that the Ward identity for the $\Delta\gamma\Delta$ vertex cannot be fulfilled
with a trimmed 3/2 propagator, which should be necessary in order to keep  the ``spin 3/2 gauge''
symmetry in the radiative case for the   $\Delta \gamma \Delta$ amplitude.
Finally, it is shown that
radiative corrections of the spin 3/2 gauge strong vertexes at one loop, reintroduce
the conventional interaction.

\bigskip


{\it Pacs}: 13.75.Gx; 14.20.Gk; 13.40.Em

\vspace{0.5in}

\end{titlepage}

\baselineskip=21pt

\newpage

\section{Introduction}

The study of massive charged spin 3/2 particles and their interactions has a great phenomenological
interest due the necessity of modelling hadron resonances, in hadron  physics experiments.
Nonetheless, the development of the theory has been plagued with difficulties and controversy. One
of them is the price to pay when we wish to fulfill Lorentz covariance.
In the Rarita-Schwinger (RS) formalism
the spin 3/2 field is given by a vector spinor $\psi^\mu$
belonging to the
$[(1/2,0)\oplus(0,1/2)]\otimes(1/2,1/2)$ representation of the Lorentz group,
enclosing both spin 3/2 and spin 1/2 fields.  In spite of the fact that on-shell we can  filter the 3/2 sector
in the equations of motion by using subsidiary conditions, thus projecting out the 1/2 one, when inverting the kinetic operator to get the propagator (an off-shell regime),  it  appears  again virtually \cite{Aurilia69}. This is not unique to the RS field: we can find other cases where the presence of several spin fields  within a representation manifest not as free particles, but as dynamical effects due to interactions \cite{Aurilia69}.

The interchange of virtual particles and its contribution to observable amplitudes is not by itself a problem
as long as the interactions do not lead to propagation of real ghosts, that is, pole contribution to the S-matrix of lower spin representation members which by construction lead to negative norm states. This is well known in Quantum Field Theory (QFT), where ghost states are not only tolerated but sometimes warmly welcome, as in the case of Faddeev-Popov Ghosts resulting from noninvariance of the measure in path integrals of nonabelian gauge symmetries \cite{WeinbergFP}.
For massive vector fields  there are ghosts  within the Gupta-Blewler quantization in the Stueckelberg's  Lagrangian \cite{Itz}. This is easy to see at the level of commutators, where  Lorentz covariance is reached by introduction of a redundant spin 0 field, rendering all components dynamical. Then, by covariance,
the commutators $[a^{\dagger \mu}, a^\nu]$ are proportional to $g^{\mu \nu}$, and the on-shell spin zero one particle states can be seen to have negative norm. For the free case, however, they are projected out by the constraint $p_\mu V^\mu = 0$.
When interactions are turned on, what needs to be warranted is that ghosts do not get physical. In the Stueckelberg's propagator used for describing the photon in the massless limit,  this is enforced for the scalar ghost since  we have electromagnetic gauge invariance (GI) and thus a coupling to a conserved current. Nevertheless, for the massive vector field in the Proca description, we  still have a virtual  spin zero contribution to amplitudes. This contribution is important, for instance in the
decay of charged pions to leptons and neutrinos mediated by virtual W vectors \cite{Aurilia69,Berme}. Observe that
the decay matrix element in spin 1 states of any helicity vanishes, so only the interchange of virtual fields
of spin 0 can account for this well known phenomenon.  It is for this reason that the spinless pion can decay through a  vector meson without violating the angular momentum conservation law.
For RS fields there are ghosts of spin 1/2 together with the on shell 3/2 contributions, they contribute virtually and do not develop a pole in the amplitude.

A more serious problem is to explicitly write down Lagrangians that lead to consistent interactions.
The  quantization of the RS free theory involves the implementation of constraints which exclude spin 1/2 states
bringing negative anticommutators into the theory, but in general the interactions change the constraints, allowing such states to re-enter again \cite{JS,AKT}. This makes the interacting theory not Lorentz
covariant, since we get superluminal solutions to the field equations,  even at the classical level \cite{Velo}.
Results have been reported for the RS field minimally
coupled to the EM field \cite{JS} and for the coupling  to a spinor and the derivative  of a scalar  field
\cite{Hagen}. This problem is related to the presence of ghost states in the free theory, but it is not simply
a consequence of virtual lower spin contributions (which are trivial to detect inspecting the propagator) but
to a subtle ``mixing'' of ghosts with the physical spectrum, resulting in a non definite spectrum.
There are good arguments to suppose that the theory can be made consistent if a complicated interaction
(not introduced minimally) is implemented \cite{Weinberg_RS,Porrati}, but  the explicit construction of such
interaction for the RS is still pending and some attempts \cite{Deser00} arrived to a negative result.

There have been many attempts to fix the form of the interaction with a pseudoscalar and a spinor (i.e., the coupling
to a $\pi$ and a $N$) on theoretical grounds. In order to keep chiral symmetry, the interaction
should involve only the  derivatives of  the pseudoscalar  field \cite{Weinberg_chiral,Peccei}.
The most general interaction term to first order in field derivatives, involves a dimensionless
parameter $Z$ \cite{Nath}. It has been argued
by Peccei \cite{Peccei} that $Z$ should be $-\frac{1}{4}$ in order to decouple the spin 1/2 states,
but \cite{Berme} shows that this decoupling does not occur. From QFT considerations,
Nath et al. \cite{Nath} (``NEK'' from now on) argued that $Z$ should be $\frac{1}{2}$ (see expression \ref{nathint} below), but later this argumentation was shown to be
incomplete \cite{Hagen},  and that for any value of $Z$ the appearance
of negative anticommutators is unavoidable. Also the inclusion of the $\Delta$ field and the consistence
of the $\pi N\Delta$ interacction has been analyzed in the framework of chiral perturbation theory
\cite{Ha05,Wies06}.
 Nevertheless, the above mentioned problems with the usual $\pi$
derivative vertex will not appear in a perturbative calculation since we  are not in presence of a classical
 background external field \cite{Weinberg_RS}, but of interactions generated by particles perturbing quantically the vacuum.

More recently Pascalutsa \cite{PASCA},  proposed a new  $\pi N \Delta$ interaction (``P'' from now on) . It also adds a derivative of the RS field, its form inspired in a ``Gauge'' transformation $\psimu \rightarrow \psimu + \partial_\mu \chi$ to which the massless RS Lagrangian is invariant. The new vertex kills
virtual 1/2 contributions in the propagator. Supporters of this interaction claim that
such decoupling is a condition of ``consistency'', since this warrants that physical amplitudes do not include the so
called ``spin 1/2 background''. But as explained above, this is not a critical
condition for consistency. The true consistency check is the absence of spin 1/2 states in the on-shell spectrum.
The necessity of the spin 1/2 background in the case of the $\Delta$ is not as evident as the case for the
spin 0 component in the $W$ boson propagator. A comparison  between conventional ${\cal L}_{NEK}$ and new ${\cal L}_P$ interactions
which should be connected by the equivalence theorem \cite{PASCA_eq},
with experimental data on elastic $\pi N$ scattering
shows that ${\cal L}_P$ cannot  reproduce them satisfactorily in contrast to ${\cal L}_{NEK}$ \cite{Nos}.

For a complete and consistent  analysis \cite{Weinberg_RS}, in addition to the strong interaction of the RS field
one should consider interaction of the $\Delta$  with radiation. Here,  the electromagnetic GI
is mandatory. Radiative $\pi N$  scattering and $\gamma N$ photoproduction through the excitation of a $\Delta$
resonance are phenomenologically relevant since they enable a determination of its magnetic dipole moment
\cite{Mariano01b,Drec01}. In order to achieve those calculations the  electromagnetic interactions were
introduced through minimal coupling, and the ${\cal L}_{NEK}$ strong coupling was adopted,
where the obtained results for the $\Delta^{++}$ and $\Delta^{+}$ where consistent.
Here, we will examine the effects of introducing minimally the electromagnetic interaction for the above mentioned
new strong interaction ${\cal L}_P$. We will see that now the 1/2 virtual propagation decoupling is not possible in  presence of the  $\Delta\gamma\Delta$
vertex, in contrast with the $\pi N$ elastic scattering. Trying to force such decoupling by inserting a spin 3/2 projector, which is inocuous in the elastic case, does not work since the $\Delta$ energy-momentum changes due to the radiated photon.
Also, it is not possible
to satisfy the Ward identity involving  the $\gamma \Delta \gamma$ vertex and $\Delta$ propagator with a trimmed
propagator keeping only the 3/2 propagation, unless one uses the mentioned projected vertexes.
We will see that after a minimal substitution in ${\cal L}_P$ spin-3/2 and electromagnetic gauge symmetries cannot coexist \cite{PASCA_report}, and that at the amplitude level we would need to throw out certain higher order terms (in a $\delta$ expansion EFT scheme described below) in order to invoke electromagnetic GI.

The paper is organized as follows. In Section \rf{recall} we review the basics of the RS field and its interactions.
In Section \rf{Gauge} we get the radiative Feynman rules for the RS theory with the new ${\cal L}_P$ interaction term
plus electromagnetic coupling introduced through minimal substitution, and discuss the spin 3/2 and electromagnetic GI consistence. In this section we also show that such
electromagnetic interactions lead to a ``spin 1/2 background''
for radiative $\pi N\rightarrow \Delta\rightarrow \pi N \gamma$ scattering, and that the introduction of a trimmed 3/2 propagator is not possible since Ward identity should be  not satisfied. In Section \rf{OneLoop}
we introduce one loop radiative vertex corrections, which  forces to reintroduce the conventional ${\cal L}_{NEK}$ coupling to
$\pi$ and $N$, so even the elastic amplitude will include a ``spin 1/2 background'' in presence of effective corrected vertexes.
Finally,  in Section \rf{Conclu}  we summarize the conclusions and discuss the implications for hadron phenomenology.

\section{RS field and its interactions}
\label{recall}

The vector-spinor $\Psi^\mu$ contains the spin 3/2 representation of the Lorentz group. The condition $p^2 = m^2$ and the requirement of keeping spin 3/2 only are equivalent to $(\slashed{p}-m)\Psi^\mu = 0$ and $\gamma_\mu \Psi^\mu = 0$. The most general first order covariant lagrangian is of the form $\bar{\Psi}^\mu \left((\slashed{p}-m) g_{\mu \nu} + A (p_\mu \gamma_\nu + p_\nu \gamma_\mu) + B\gamma_\mu \slashed{p} \gamma_\nu + mC \gamma_\mu \gamma_\nu \right)\Psi^\nu$, and imposing the condition that the equations of motion are equivalent to  $(\slashed{p}-m)\Psi^\mu = 0$ and $\gamma_\mu \Psi^\mu = 0$ we get $B = \frac{3}{2}A^2+A+\frac{1}{2}$, $C = 3A^2 + 3A +1$ and $A \neq -\frac{1}{2}$. Thus, a family of infinite equivalent lagrangians and equations of motion is obtained.

The equations of motion for a spin 3/2 field $\psi^\mu$ are thus the Dirac equation for each component
(enforcing the relativistic dispersion relation and definite parity) plus the constraints that select the
``pure spin 3/2'' representation,  $\gnp\psinu=0$ (The condition $\partial^\nu \psinu=0$, which is also required to project out all spin 1/2 components, follows from the former and the Dirac equation on each component). These field equations are invariant under the so called ``contact transformation''.
\br \psi^\mu \rightarrow \psi'^\mu = R(a)^{\mu \nu} \psinu \equiv
(\gmmunnu + a \gmp \gnp) \psinu\label{contact} ,\er
 \cite{ElAmiri,Nos}, which is related with the abovementioned existence of constraints, since it changes only the spin 1/2 sector that is projected out by them.
This is not privative of the 3/2 field: the vector representation within the $(1/2,1/2)$ space obeys the Proca equation for a field of mass $\mu$  with the subsidiary condition $\partial \cdot A=0$ and  it is possible
to make the transformation $A^\rho \rightarrow A^\rho +\lambda/\mu^2 \partial^\rho(\partial \cdot A)=
(g^{\rho}_\alpha +\lambda/\mu^2 \partial^\rho \partial_\alpha)A^\alpha$, which only affects the spin zero sector, leading to equivalent equations of motion and different propagators \cite{Itz}.

The Lagrangian can be expressed as
\cite{Nos}
\br \L_{free} &=&  \barpsimux \K(\partial,A)^{\mu \nu}\psi_\nu(x),\label{freelag2}\er
where
\br \K(\partial,A)^{\mu \nu}&=&
R\left(-\onehalf(1+A)\right)^{\mu \mu'}\left\{\epsilon_{\mu' \nu' \alpha \beta}\partial
^\alpha\gamma^\beta \g_5 + i m \sigma_{\mu' \nu'}\right\}R\left(-\onehalf(1+A)\right)^{\nu' \nu} \label{freelag2b}\er
To express it in momentum space we make the replacement $i \partial_\mu = i{\partial \over \partial x^\mu} \rightarrow p_\mu$.
Equations \rf{freelag2b} and \rf{freelag2} are equivalent to the conventionally set in most of the literature
(see for instance \cite{Berme,Nath})\footnote{Expressing the Lagrangian with the aid of the matrices $R$ makes it much easier to manipulate computations involving the
contact invariance.}.
The spin-$\threehalf$ propagator $G(p,A)_{\beta\nu}$ should satisfy
\br  \K(p,A)_{\mu}^{\beta}G(p,A)_{\beta\nu}=g_{\mu
\nu},\label{propdef} \er
for any value of $A$, using properties of $R$ \cite{Nos} and the Eq. \rf{freelag2b} can be written conveniently as
\br
G(p,A)^{\mu \nu}&=&
\left[R^{-1}\left(-\onehalf (1+A)\right)^\mu_\alpha\right]
G\left(p,-1\right)^{\alpha \beta}\left[R^{-1}\left(-\onehalf (1+A)\right)_{\beta}^\nu\right],\label{propt1}
\er
where $G(p,-1)$ is the well known propagator for $A=-1$:
\br
G\left(p,-1\right)_{\mu \nu}
&=&-\left[{ \ps + m\over p^2 - m^2}\hat{P}^{3/2}_{\mu \nu} -
{2\over 3m^2}(\ps + m)(\hat{P}^{1/2}_{22})_{\mu \nu} +{1\over\sqrt{3} m} (\hat{P}^{1/2}_{12}+\hat{P}^{1/2}_{21})_{\mu
\nu}\right].\nn\\&&\label{propproj}
\er
We have introduced $P^k_{ij}$  which projects on the $k=3/2,1/2$ sector of the representation space,
with $i,j=1,2$ indicating the subsectors of the $1/2$ subspace, and are defined as
\br (\hat{P}^{3/2})_{\mu \nu}&=&g_{\mu \nu}-{ 1 \over
3}\gamma_\mu\gamma_\nu -{1\over 3p^2}\left[\pslash \g_\mu p_\nu +
p_\mu\g_\nu\pslash\right],\nonumber\\
(\hat{P}^{1/2}_{22})_{\mu \nu}&=&{p_\mu p_\nu \over p^2},\nonumber\\
(\hat{P}^{1/2}_{11})_{\mu \nu}&=& \gmunu - \hat{P}^{3/2}_{\mu \nu}
-(\hat{P}^{1/2}_{22})_{\mu \nu},\nn\\
&=&
(g_{\mu \alpha}-{p_\mu  p_\alpha\over
p^2})(1/3\gamma^\alpha \gamma^\beta)(g_{\beta \nu}-{p_\beta p_\nu \over
P^2}),\nonumber\\
(\hat{P}^{1/2}_{12})_{\mu \nu}&=&{1\over \sqrt{3}p^2}(p_\mu p_\nu
-\ps
\gamma_\mu p_\nu),\nonumber\\
(\hat{P}^{1/2}_{21})_{\mu \nu}&=&{1\over \sqrt{3}p^2}(-p_\mu
p_\nu+\ps p_\mu \gamma_\nu ).\label{spinprojectors} \er
As can be seen from \rf{propproj}, the spin 1/2 sector does not develop a pole and only manifests
as virtual states.
When vertex interactions are present, and as amplitudes should not depend on $A$, we fix
their A-dependence by demanding point invariance of the interaction terms \cite{Nos}. However, still getting
A-independent amplitudes
we have several criteria to fix the interaction vertexes. We will analyze the elastic and radiative $\pi N$
scattering amplitude, and we begin
describing the strong $\pi \Delta N$ vertex.
In \cite{Nath} the leading order coupling between a $\Delta$ , $N$ and $\pi$ is studied
based on the nonlinear realization of the chiral symmetry, imposing conditions of contact invariance.
They get the interaction
\begin{equation}
 {\cal L}_{NEK}= {f_{\pi N\Delta}\over m_\pi} \bar{\psi} \partial_\mu \phi^\dag\cdot {\bf T} R\left(\onehalf (1+4Z)A+Z\right)^{\mu \nu} \Psi_\nu + h.c.,
\label{nathint}
\end{equation}
where ${\bf T}$ is the
$\Delta \rightarrow  N$ isospin transition operator. Since the matrix $R(\onehalf (1+4Z)A+Z)$ can be written as $R(-\onehalf(A+1))R(-Z-\onehalf)$,
the contraction of such vertexes with the propagator $G^{\alpha
\beta}(p,A)$ given in Eq.\rf{propt1} leads to A-independent $\pi N$  scattering amplitudes.
The value $Z$ was then fixed to $\onehalf$ from field theoretic arguments.
Much more recently, in \cite{PASCA} and \cite{Brief} a higher order interaction\footnote{
In \cite{PASCA} it is introduced the interaction when $A=-1$, but in ref.\cite{Brief} the expression is generalized.}
\begin{equation}
{\cal L}_{P} = {f_{\pi N\Delta}\over m_\pi m} \; \bar{\psi} \partial_\mu \phi^\dag\cdot {\bf T}
\epsilon^{\mu \sigma \rho \nu }\gamma_{\nu} \gamma_5 \partial_\rho
 R\left(- \frac{1}{2}(A+1)\right)_{\sigma}^\eta
\Psi_\eta + h.c.,
\label{pascaint}
\end{equation}
was proposed where, according to the notation above, the value $Z=-1/2$ was adopted.
The basis for the proposal  was the symmetry for ${\cal L}_{free}(m=0)$ (see Eqs.\rf{freelag2} and
\rf{freelag2b}) under a ``spin 3/2 gauge-like'' transformation $\Psi_\mu \rightarrow \Psi_\mu + \partial_\mu \Phi$ (where $\Phi$ is a spinor)  of
the RS field \cite{PASCA_report}, which leads to introducing  the dependence $
\epsilon^{\mu \sigma \rho \nu }\gamma_{\nu} \gamma_5 \partial_\rho \Psi_\sigma$ for the Lagrangian \rf{pascaint}, similar to that in ${\cal L}_{free}(m=0)$ in Eq.\rf{freelag2b}.
  Then, for the $\pi N \rightarrow \Delta \rightarrow \pi N$ the amplitude resulting from the interaction vertex \rf{pascaint} has the following property ($\Gamma^{\mu \sigma}(p)\equiv
   \epsilon^{\mu \sigma \rho \nu }\gamma_{\nu} \gamma_5 p_\rho$ and isospin factors ommited)
\begin{equation}
 \left({f_{\pi N\Delta}\over m_\pi m}\right)^2\Gamma_{\mu \rho}(p) G^{\rho \sigma}(p,-1) \Gamma_{\sigma \tau}(p) = -\left({f_{\pi N\Delta}\over m_\pi }\right)^2\frac{p^2}{m^2}{ \ps + m\over p^2 - m^2} P^{(3/2)}_{\mu \tau}(p),
\label{projprop}
\ee
since $G(p,-1)$ is built with $P_{22}^{\onehalf},P_{12}^{\onehalf},P_{21}^{\onehalf}$ from \rf{spinprojectors},
satisfying
\br
P_{22}^{\onehalf \mu\sigma}\Gamma_{\sigma\nu }(p)&=&0,\nn \\ \Gamma_{\sigma \mu }(p)P_{22}^{\onehalf \mu \sigma}&=&0,\nn \\
P_{12}^{\onehalf \mu\sigma}\Gamma_{\sigma\nu }(p)&=&0,\nn \\ \Gamma_{\mu \sigma}(p) P_{21}^{\onehalf \sigma\nu}&=&0,\label{proyp}\er
due to the  ``transversality property'' of the spin 3/2 gauge invariant vertex
\begin{equation}
\Gamma^{\mu \nu}(p)p_\nu= \Gamma^{\mu \nu}(p)p_\mu = 0,
\label{transversality}
\end{equation}
and the constraint conditions
$\gamma^\alpha, p^\alpha P^{(3/2)}_{\alpha \beta}(p)=P^{(3/2)}_{\alpha \beta}(p)\gamma^\beta, p^\beta=0$.
This shows that the spin 1/2 states carried by  the
propagator \rf{propproj} do not contribute to the elastic scattering.
We mention that it is possible to pass form $\L_{free}+\L_{NEK}$ to $\L_{free}+\L_{P}+\L_{C}$, where  $\L_{C}$ represents  contact terms not involving the $\Delta$ field, by performing a linear transformation $\psi^\mu \rightarrow
\psi^\mu - {f_{\pi N\Delta}\over m_\pi m} \psi \partial_\mu \Phi\cdot {\bf T}^\dag$. It was argued that when we build for example
the $\pi N$ scattering amplitude the contribution coming from $\L_{C}$ could be hidden within the non resonant background contributions \cite{PASCA_eq} , nevertheless it has been shown that calculations with the original and transformed Lagragians are not equivalent \cite{Nos}. Observe that in trying to eliminate spin 1/2 backgrounds, the interaction ${\cal L}_P$ introduces a $p^2$ factor in the amplitude \rf{projprop}, increasing the resonance contribution away from $m^2$.

Finally notice that (using $[\pslash,P^{3/2}]=0$) if we make the replacement
 \br\Gamma_{\sigma \tau}(p) &\rightarrow&  \tilde{\Gamma}_{\sigma \tau}(p)=P^{(3/2)\sigma}_{\mu}
(p) \Gamma_{\sigma \tau}(p),\nn\\
G\left(p,-1\right)_{\mu \nu}&\rightarrow&\tilde{G}\left(p,-1\right)_{\mu \nu}
=-{ \ps + m\over p^2 - m^2}\hat{P}^{3/2}_{\mu \nu},\label{p3_2}
\er
in the left hand side of \rf{projprop}, we do not alter the result.
Nevertheless we will see in the next section that this is not true in the radiative case and that the trimmed propagator \rf{p3_2} cannot fulfill the Ward identity with the $\gamma \Delta \gamma$ vertex and furthermore, it has not inverse.

\section{The spin 3/2 and electromagnetic gauge coexistence}
\label{Gauge}

In both ${\cal L}_{free}$ and ${\cal L}_P$ appear derivatives, the  electromagnetic GI
will be fulfilled if the coupling to photons is done through minimal substitution $\partial_\mu \rightarrow \partial_\mu - i q A_\mu$, where $A_\mu$ is the electromagnetic field.
We do not consider nonminimal couplings here, since we are
studying the coexistence between both the ``spin 3/2 '' (eq.\rf{transversality})  and the electromagnetic gauge invariances.
The study of this point will be equivalent to analyzing the invariance of the Lagrangian under  $\Psi_\mu \rightarrow \Psi_\mu + \partial_\mu \Phi$
in the radiative case.
Making minimal substitutions in ${\cal L}_{free}+{\cal L}_P$, we get the electromagnetic interaction
Lagrangians ($q_\Delta=2e,e,-e$ for $\Delta^{++,+,-}$ and
$q_\pi=e,-e$ for $\pi^{+,-}$)
\begin{eqnarray}
{\cal L}_{\Delta \gamma \Delta} &=& i q_\Delta  \bar{\Psi}^\mu \Gamma(A)_{\mu\sigma} \Psi^\sigma\nn\\
{\cal L}_{\Delta N \gamma \pi}^{(1)}  &=& i{f_{\pi N\Delta}\over m_\pi m}
q_\pi \bar{\psi}(\phi^\dag\times{\bf T})_3
\Gamma(A)_{\nu\mu}(\partial^\mu \Psi^\nu)   + h.c.\nn\\
{\cal L}_{\Delta N \gamma \pi}^{(2)}  &=& i{f_{\pi N\Delta}\over m_\pi m}
q_\Delta \bar{\psi}(\partial^\beta \phi^\dag\cdot{\bf T})\Gamma(A)_{\beta\nu}\Psi^\nu   + h.c..\label{radiat}
\end{eqnarray}
${\cal L}_{\Delta \gamma \Delta}$ arises from minimal substitution in ${\cal L}_{free}$,
while ${\cal L}_{\Delta N \gamma \pi}^{(1)}$
and ${\cal L}_{\Delta N \gamma \pi}^{(2)}$ arise from substitution of pion and $\Delta$ derivatives, respectively, in ${\cal L}_P$.  We have dropped R matrices since they cancel up in building the amplitudes.

It is clear that ${\cal L}_{\Delta \gamma \Delta}$, and ${\cal L}_{\Delta N \gamma \pi}^{(2)}$
defined in Eq.\rf{radiat}, are no longer spin 3/2 GI since the $\Delta$ derivative was replaced. We could try to make a linear transformation
\br{\bar \psi}_\rho \rightarrow {\bar \psi}_\rho - {f_{\pi N\Delta}\over m_\pi m}
q_\Delta \bar{\psi}(\partial^\beta \phi^\dag\cdot{\bf T})\Gamma(A)_{\beta\rho}\sigma^{-1\rho\nu}\nn\er
in ${\cal L}_{free}+{\cal L}_P+{\cal L}_{\Delta N \gamma \pi}^{(1)}+{\cal L}_{\Delta N \gamma \pi}^{(2)}$ to get
${\cal L}_{free}+{\cal L}_P+{\cal L}_{\Delta \gamma \Delta}+{\cal L}_{\Delta N \gamma \pi}^{(1)}+\tilde{{\cal L}}_{\Delta N \gamma \pi}^{(2)}+{\cal L}_{C}+{\cal L}_{A^2}$, where now $\tilde{{\cal L}}_{\Delta N \gamma \pi}^{(2)}$ is spin 3/2 GI. Nevertheless the total Lagrangian will not posess electromagnetic GI  since the term coming from minimal substitution in the $\partial_\mu\psi^\nu$ in ${\cal L}_P$ is not present any more after the transformation.
On the other hand ${\cal L}_{\Delta \gamma \Delta}$ still violates spin 3/2 GI, and only a nonlinear transformation containing the $\Delta$ field should be applied to eliminate it but at the price of generating an infinite number of supplementary terms \cite{PASCA_eq}. Then, we arrive at the conclusion that it is not possible to fulfill both spin 3/2 and electromagnetic GI symmetries in a closed way, we can only hope to  fulfill them by making 
some approximations.

One possible framework to do this is the chiral perturbation theory ($\chi PT$) including $\pi$, $N$ and $\Delta$
degrees of freedom
which is attractive in that it is supposed to be a low-energy expansion of QCD. Here it is possible to compute the pion mass dependence of static quantities, such as nucleon mass, magnetic moments and momentum dependence of $\pi\pi$ and $\pi N$ scattering process . There are here two light scales and a parameter such that $\delta \equiv (m - m_N)/(\Lambda_{\chi PT}= \mbox{ 1GeV})$  and $m_\pi
/\Lambda_{\chi PT} \sim \delta^2 $,
in terms of which we can define an order $\delta^n$ for a given amplitude through a power counting recipe \cite{PASCA_report} (we need to replace $f_{\pi N\Delta}/m_\pi\simeq 0.014 MeV^{-1}  \rightarrow h_A/2(f_\pi = 92MeV) \simeq 0.016MeV^{-1}$).  An exchanged momentum $q$ (through $\pi$ or $\gamma$) counts as $\delta$ or $\delta^2$
if it is close to $m - m_N$ or $m_\pi$ regions respectively.
$n$ depends on  the vertex Lagrangian order defined as the sum of de number of $\pi$ field derivatives, plus $A_\mu$
field derivatives, plus the order of the charge $e$. In addition we have a count for the $\Delta$-propagators when $q\sim m - m_N$ that differentiates
the case when  $G(p,-1)$ participates in the amplitude through an s-pole channel subgraph (one-$\Delta$-reducible
graph)  or in another way.
Without going into details about this counting  it is clear that the spin 3/2 gauge invariant ${\cal L}_{\Delta N \gamma \pi}^{(1)}$ gives a contribution of order $\delta$, while ${\cal L}_{\Delta N \gamma \pi}^{(2)}$ of order $\delta^2$ since when we have $\partial_\mu\psi^\nu/m$ in ${\cal L}_{P}$ it does not contribute to the counting\footnote{If $p=p_N + q$ then $p/\Lambda_{\chi PT} =p_N/\Lambda_{\chi PT}
+ q/\Lambda_{\chi PT}\sim 1 + \delta^2  \simeq 1$ at threshold}, but after the
replacement  $\partial_\mu \rightarrow \partial_\mu - i q A_\mu$ it acquires order $\delta^2$.
Also it is clear that in a given $\delta^n$ order amplitude where ${\cal L}_{\Delta N \gamma \pi}^{(1)}$ accounts for the $\Delta \pi N \gamma$  vertex, ${\cal L}_{\Delta N \gamma \pi}^{(2)}$ should contribute at order
$\delta^{n+1}$ and one could throw it out (at the price of not fulfilling electromagnetic GI) to get spin 3/2 symmetry.

Still we have another question referring to the Ward identity necessary to account for the electromagnetic GI of the amplitude in presence of the $\Delta \gamma \Delta$ vertex. This identity reads
\br
i  G(p',-1)^{\mu \alpha} \Gamma_{\alpha \beta}(p-p')  G(p,-1)^{\beta \nu}= G(p,-1)^{\mu \nu} -
G(p',-1)^{\mu \nu},\label{Wardi}\er
and it can be demonstrated that the full propagator \rf{propproj} satisfies it. In the case of radiative
$\pi N$ scattering the contribution of the $\Delta \gamma \Delta$ vertex to the amplitude is
\br
\M_{\Delta \gamma \Delta} &=& i \left({f_{\pi N\Delta}\over m_\pi m}\right)^2q_
q'^\mu \Gamma_{\mu \sigma}(p')
 G(p',-1)^{\sigma \alpha} \Gamma_{\alpha \beta}(e*)  G(p,-1)^{\beta \rho}\Gamma_{\rho \nu}(p)q^\nu,\label{deltagamadelta}
\er
where $e_\mu$ is the photon polarization vector
Note that now the property \rf{projprop} cannot be fulfilled neither by the first pair of vertexes in
 \rf{deltagamadelta} nor by the second one,  since the middle vertex is evaluated at $e^*$
and we lost the condition of transversality, that is

\begin{equation}
\Gamma^{\mu \nu}(e*)p_\nu= \Gamma^{\mu \nu}(e*)p_\mu \neq 0,
\label{nontransversality}
\end{equation}
rendering \rf{proyp}
no longer valid, and then the replacement proposed \rf{p3_2} is not possible in the radiative case. Of course, all these is consequence of the fact that
${\cal L}_{\Delta \gamma \Delta}$ does not posess spin 3/2 GI. Then we conclude that in any radiative amplitude involving the
radiation from the $\Delta$ the coupling to virtual spin 1/2 states is unavoidable.
Some treatments based on $\chi$PT \cite{PASCA_report}, in spite of the above observation, assume
$\tilde{G}\left(p,-1\right)_{\mu \nu}$ as the propagator in the radiative amplitude \rf{deltagamadelta} adopting
$\Gamma \sim \tilde{\Gamma}$. Within this replacement the  electromagnetic vertex is reduced to  $\Gamma_{\alpha \beta}(e^*)\simeq g_{\alpha \beta} \eeslash$ since the on-shell constraints $\gamma^ \mu P^{3/2}_{\mu\nu}=p^ \mu P^{3/2}_{\mu\nu}=0$ are enforced. But it is not possible to think that \rf{Wardi} is satisfied as an identity by $\tilde{G}$ and $g_{\alpha \beta} (\eeslash\rightarrow \pslash -\ppslash)$, it is only valid when we sandwich \rf{Wardi} between  $\tilde{\Gamma}$ which we have seen is not right in the radiative case.
 Finally it is important to mention that in the same reference the equivalent
identity (for the dressed vertex and propagator)
\br
i  \Gamma_{dressed}^{\mu \beta}(p-p') &=&G_{dressed}^{-1}(p',-1)^{\mu \nu}-  G_{dressed}^{-1}(p,-1)^{\mu \nu}\nn\\
&=&G^{-1}(p',-1)^{\mu \nu} -  G^{-1}(p,-1)^{\mu \nu} + \Sigma(p',-1)^{\mu \nu}-\Sigma(p,-1)^{\mu \nu}
,\label{Wardi2}
\er
where $\Sigma$ is the one loop pion-nucleon $\Delta$ self energy, is used to fix $F(0)=1 -\Sigma '(m)$ in $\Gamma_{d \mu \beta}(q)=F(q^2)\Gamma_{\mu \beta}(q)\approx F(q^2) g_{\alpha \beta} \qslash$ with $G \sim \tilde{G}$, nevertheless $\tilde{G}$ is not invertible and the spin 1/2 like  assumption $\tilde{G}^{-1}=\pslash - m$  (necessary to get the result) is  wrong. Again we interpret that a projected identity $P^{3/2}(p') \rightarrow \cdot \cdot \cdot \leftarrow  P^{3/2}(p)$ is {\it assumed} in place of \rf{Wardi2} since
$P^{3/2}(p') \Gamma_{\mu \beta} (e^*)= \Gamma_{\mu \beta} P^{3/2}(e)\neq 0$  .

We can see how the $\delta$ expansion works to get an approximated electromagnetic gauge invariant amplitude in the case of
$\pi^0 p$ radiative scattering amplitude (the simplest one) as shown Figure \ref{pi0p}.
\begin{figure}
  \begin{center}
    \includegraphics[width=5cm]{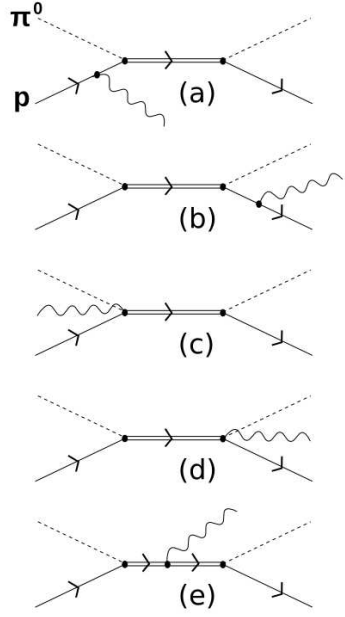}
  \end{center}
  \caption{\small  Feynmann graphs corresponding to the radiativer  $\pi^0p$ scattering.}
  \label{pi0p}.
\end{figure}
Note that if we want to keep spin 3/2 GI at leading order in the $\delta$ expansion the Lagrangian
on the contributions of ${\cal L}_{T} \equiv {\cal L}_{free} + {\cal L}_P + {\cal L}_{\Delta \gamma \Delta} +
{\cal L}_{\Delta N \gamma \pi}^{(1)}$ should be considered since
${\cal L}_{\Delta N \gamma \pi}^{(2)}$(which violates spin 3/2 GI)  is of next order regards ${\cal L}_{\Delta N \gamma \pi}^{(1)}$. Since the isospin operator in this last lagrangian is $(\phi^{0\dag}\times{\bf T})_3=0$,
we have no contributions from the figures 1(c) and 1(d), that should contribute if we consider
${\cal L}_{\Delta N \gamma \pi}^{(2)}$ since the isospin operator is $\phi^{0\dag}\cdot{\bf T}=T_3$
which will give a contribution between the $\Delta^+$ and $p$. It can be shown by using the Feynman rules
obtained from ${\cal L}_{T}$, the Ward identity \rf{Wardi} and making the replacement $e^*\rightarrow k$, in order to corroborate the electromagnetic GI that
\br
{\cal M}_{T}(e^* \rightarrow k) = - {\cal M}^{(2)}_{c+d}(e^*\rightarrow k),\label{gi2}
\er
where in the right hand side we have the amplitude corresponding to the graphs 1(c)+1(d)  calculated with the
excluded ${\cal L}_{\Delta N \gamma \pi}^{(2)}$ Lagrangian. Then the electromagnetic GI (${\cal M}_{T}(e^* \rightarrow k) =0$) is not fulfilled, we only get an approximation if we drop
${\cal M}^{(2)}_{c+d}(e^*\rightarrow k)$ since it is of higher order in the $\delta$ expansion.
Similar results would be obtained for other different processes as for example Compton scattering
\cite{PascaVan}. On the other hand if we use
${\cal L}_{T} \equiv {\cal L}_{free} + {\cal L}_{NEK} +
{\cal L}_{\Delta \gamma \Delta} + {\cal L}_{NEK}(\partial_\mu \rightarrow q_\pi A_\mu)$ then
elctromagnetic GI is perfectly fulfilled.

\section{One loop radiative corrections}
\label{OneLoop}

Once we have established the spin 3/2 and electromagnetic GI can coexist only approximately since in the radiative case the property \rf{transversality} is spoiled, it is natural
to suspect that one loop radiative corrections to $\pi N$  scattering through the $\Delta$ (bubbles in Figure \ref{exact})
may also spoil that property. To check it, lets study  these corrections to the $\pi N \Delta$
vertex that are explicitly shown in Figure \ref{oneloop}.
Since this is a nonrenormalizable theory, we expect that it will be needed to include higher order counterterms to cancel all divergences. Recall that according to the modern conception of renormalizability
all theories are renormalizable in the sense that their infinities can be absorbed by appropiate counterterms,
but a systematic loop expansion for nonrenormalizable theories is useless (there is not any predictive power)
since at each stage new arbitrary parameters must be included \cite{Weinberg2}. So, there is no point in calculating
loop corrections in all detail. It is relevant, however, to know the form of the terms with coupling constant
of lowest inverse mass dimmension, since at low energy they are the only relevant.
In EFT we must include {\it all} interaction terms compatible with the
symmetries of the theory \cite{Weinberg2} . Nevertheless, for ${\cal L}_{P}$ in \rf{pascaint} the lower derivative term
is absent, and is the one  included by ${\cal L}_{NEK}$  in Eq. \rf{nathint}.
As we will see, this term emerge in the one-loop corrected vertex. We will treat divergent integrals as in dimmensional
regularization. We will not isolate the divergence and perform the regularization because we are just interested in
finding the form of the more relevant counterterms at low energies.

\begin{figure}
  \begin{center}
    \includegraphics[width=8cm]{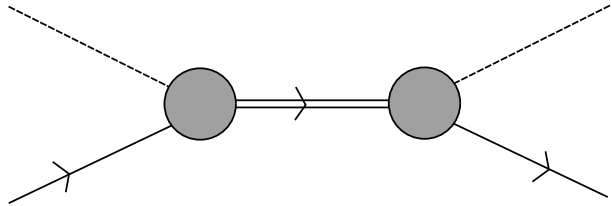}
  \end{center}
  \caption{\small $\Delta$-pole contribution to $\pi N$ scattering with radiative corrected vertexes.}
  \label{exact}
\end{figure}

\begin{figure}
  \begin{center}
    \includegraphics[width=12cm]{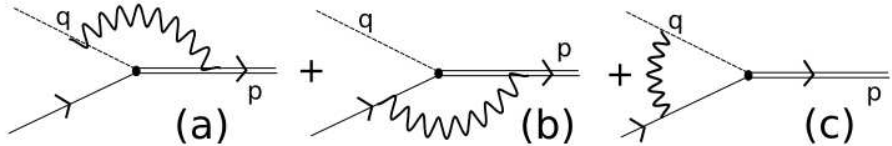}
  \end{center}
  \caption{\small One loop radiative corrections to the $\pi N$ vertex.}
  \label{oneloop}
\end{figure}

Due to the circulation of loop momentum, not all terms exhibit a $q^1 p^0$ contribution in their Taylor expansions.
Let us take for instance the vertex in diagram \ref{oneloop}(a) which reads
\begin{eqnarray}
&& \displaystyle q_\pi q_\Delta {f_{\pi N\Delta}\over m_\pi m}\; \int \frac{d^4s}{(2\pi)^4} (2q^\mu -s^\mu)  \frac{g_{\mu \nu}}{s^2}
          \frac{1}{\left[(q-s)^2-m_\pi^2 \right]} \frac{(\slashed{s}+m)}{\left[(p-s)^2-m^2 \right]}
          \epsilon^{\delta \sigma \rho \tau} \gamma^5 \gamma_\sigma \nonumber \\
&\times& \displaystyle(p_\delta - s_\delta)(q_\tau - s_\tau) \left( g_{\rho \eta} - \frac{1}{3} \gamma_\rho \gamma_\eta -\frac{1}{3m} \gamma_\rho (p_\eta - s_\eta)
             \right) \epsilon^{\nu \eta \alpha \beta} \gamma^5 \gamma_\alpha,
\label{diagA}
\end{eqnarray}
where the Eqs. \rf{propproj}, \rf{spinprojectors}and \rf{transversality} were used.  The diagram \ref{oneloop}(b) will contribute as

\begin{eqnarray}
& \displaystyle q_N q_\Delta {f_{\pi N\Delta}\over m_\pi m}\; \int \frac{d^4s}{(2\pi)^4} \frac{g_{\mu \nu}}{s^2}
          \gamma^\mu \frac{(\slashed{p}-\slashed{q}-\slashed{s}+m_N)}{\left[(p-q-s)^2-m_N^2 \right]}
          \frac{(\slashed{s}+m)}{\left[(p-s)^2-m^2 \right]}
          \epsilon^{\delta \sigma \rho \tau} \gamma^5 \gamma_\sigma  \times \nonumber \\
& \displaystyle (p_\delta - s_\delta)
          q_\tau \left( g_{\rho \eta} - \frac{1}{3} \gamma_\rho \gamma_\eta -\frac{1}{3m}  \gamma_\rho (p_\eta - s_\eta)
               \right)
       \epsilon^{\nu \eta \alpha \beta} \gamma^5 \gamma_\alpha,
\label{diagB}
\end{eqnarray}
where the anomalous nucleon magnetic moment contribution was omitted for simplicity.
The diagram \ref{oneloop}(c) is not considered since it is not possible to get
a term of order $q^1 p^0$ from it.
Taylor-expansions of the above vertexes around $p=0$, $q=0$ lead
to series of interaction terms to all orders in $q$ and $p$ (i.e. $\pi$ and $\Delta$ derivatives), which dress the bare vertexes.
The $q^0 p^1$ contribution in \rf{diagA}  (the corresponding in \rf{diagB} vanishes) reads
\br
&&q_\pi q_\Delta {f_{\pi N\Delta}\over m_\pi m}\; p_\delta \int \frac{d^4s}{(2\pi)^4} s_\nu s_\tau \frac{g_{\mu \nu}}{s^2}
          \frac{1}{\left[s^2-m_\pi^2 \right]}
          \frac{(\slashed{s}+m)}{\left[s^2-m^2 \right]}
          \epsilon^{\delta \sigma \rho \tau} \gamma^5 \gamma_\sigma
          \left( g_{\rho \eta} - \frac{1}{3} \gamma_\rho \gamma_\eta\right)\nn\\
&\times&\epsilon^{\nu \eta \alpha \beta} \gamma^5 \gamma_\alpha,
\er
being divergent and non derivative in the $\pi$ field. It
breaks chiral symmetry, and so a chiral symmetry breaking counterterm is required, which is not an
inconsistency since chiral symmetry is only approximate. After renormalization,
this contribution should be kept small at low energies. We have a contribution $q^1 p^0$ from \rf{diagA}, being
\br
&&q_\tau A^{\tau \beta} \equiv q_N q_\Delta {f_{\pi N\Delta}\over m_\pi m}\nn\\&\times&  q_\tau \int \frac{d^4s}{(2\pi)^4} s_\nu s_\delta \frac{g_{\mu \nu}}{s^2}
          \frac{(m_N-\slashed{s})}{\left[s^2-m_N^2 \right]}
          \frac{1}{\left[s^2-m_\pi^2 \right]}
          \epsilon^{\delta \sigma \rho \tau} \gamma^5 \gamma_\sigma
          \left( g_{\rho \eta} - \frac{1}{3} \gamma_\rho \gamma_\eta\right)
       \epsilon^{\nu \eta \alpha \beta} \gamma^5 \gamma_\alpha\nn\\
\label{NathContribA}
\er
while  from \rf{diagB} one

\br
&&q_\tau B^{\tau \beta}\equiv q_N q_\Delta {f_{\pi N\Delta}\over m_\pi m} \nn\\ &\times& q_\tau \int \frac{d^4s}{(2\pi)^4} s_\nu s_\delta \frac{g_{\mu \nu}}{s^2}
          \gamma^\mu \frac{(m_N-\slashed{s})}{\left[s^2-m_N^2 \right]}
          \frac{(\slashed{s}+m)}{\left[s^2-m^2 \right]}
          \epsilon^{\delta \sigma \rho \tau} \gamma^5 \gamma_\sigma
          \left( g_{\rho \eta} - \frac{1}{3} \gamma_\rho \gamma_\eta\right)
       \epsilon^{\nu \eta \alpha \beta} \gamma^5 \gamma_\alpha.
\nn\\\label{NathContribB}
\er
Then, we will face (dropping terms in odd powers of $s$) with infinite integrals
\begin{equation}
G^1_{\nu \delta}(q) = \int \frac{d^4s}{(2\pi)^4} s_\nu s_\delta \frac{1}{s^2}
          \frac{m_N}{\left[s^2-m_N^2 \right]}
          \frac{m}{\left[s^2-m^2 \right]}
\end{equation}
\begin{equation}
G^2_{\nu \delta}(q) = \int \frac{d^4s}{(2\pi)^4} s_\nu s_\delta
          \frac{1}{\left[s^2-m_N^2 \right]}
          \frac{1}{\left[s^2-m^2 \right]}
\end{equation}
\begin{equation}
G^3_{\nu \delta}(q) = \int \frac{d^4s}{(2\pi)^4} s_\nu s_\delta
          \frac{1}{\left[s^2-m_\pi^2 \right]}
          \frac{1}{\left[s^2-m^2 \right]},
\end{equation}
which after performing the angular integration can be written as $g_{\nu \delta} \times \mbox{divergent integral}$. So, we have that

\be
q_\tau A,B^{\tau \beta} = q_\tau C^{1,0}(g^{\tau \beta} + Z^{1,0} \gamma^\tau \gamma^\beta),
\ee
where $C^{1,0}$ and $Z^{1,0}$ are infinite constants. The indexes  indicate that they are effective coupling constants
corresponding\footnote{In this languaje, for instance,
$C^{1,1}$ would correspond to the ${\cal L}_P$ vertex  coupling constant, and $Z^{1,1}=-\frac{1}{2}$} to $q^1 p^0$.
In order to cancel
infinities that go as $q^1 p^0$, then we must add to the original Lagrangian \rf{pascaint} an interaction of the form \rf{nathint} with infinite bare constants.

\section{Concluding remarks}
\label{Conclu}

A QFT involving effective Lagrangians for a   spin 3/2 particles is nonrenormalizable, and so it must include
all order interaction terms compatible with its symmetries.
In this sense, there is no fundamental reason
for suppressing lowest order derivative interaction ${\cal L}_{NEK}$ in Eq. \rf{nathint} and considering the second order
${\cal L}_{P}$ in Eq.\rf{pascaint} as the leading term
at low momenta. The reason for using this Lagrangian  was a symmetry for ${\cal L}_{free}(m=0)$ under a
gauge-like transformation of the RS field, where the corresponding strong $\pi N \Delta$ vertex fulfils the relation \rf{transversality} on which resides \rf{projprop}, and so  the avoiding of virtual 1/2 states contributions in the elastic scattering amplitude. Spin 3/2 GI is not present in the
theory, since $m>0$, but the mass term does not spoil the property \rf{transversality}. Instead, EM interactions introduced through a minimal substitution  spoil property \rf{transversality} in the radiative vertex,
reintroducing a spin 1/2 background.
In order to enforce spin 3/2 GI in Lagrangians which are initially noninvariant, in Ref. \cite{PASCA_eq}
was proposed to make linear transformations in the RS field in a manner that the original Lagrangian
is transformed into a gauge invariant one plus certain contact contributions that do not involve the RS field. These last contributions
are supposed to be absorbed in other non resonant background contributions, what was shown not to be true \cite{Nos}. The new Lagrangian fulfilling spin 3/2  GI  has no more electromagnetic GI since it contains a new $\delta$-derivative term, and a new minimal substitution brings it into GI breaking again the spin 3/2 symmetry.
In order to go out from this loop  a $\delta$ perturbative expansion was proposed \cite{PASCA_report} within a $\chi_{PT}$ framework,
which is supposed to enable us to speak of coexistence between different gauge symmetries at a given order but not in a closed way.
The problem with that particular proposal is that by changing the Lagrangians to get the spin 3/2 GI,
we are not in presence of the perturbative expansion of the
original theory since the original Lagrangians were changed. In addition the dropped terms to get the mentioned coexistence are not in general small. This procedure would make sense only for treating specific reactions in the low energy region for photons($\gamma$), but is not useful in general and specially for the vector current in neutrino induced processes where the energies carried by the weak bosons are not small compared to the $\delta$ scale.

We show that the Ward identity for the $\Delta\gamma\Delta$ vertex, which cannot be turned into a spin 3/2 gauge invariant one through a linear transformation, cannot be fulfilled
with a non invertible trimmed 3/2 propagator, which appears when we use the projected vertexes.

Observe that the occurrence of such spin 1/2 background is not by itself a problem, since as exposed in the introduction
the interchange of virtual unphysical particles, usually contribute to amplitudes
and sometimes that contribution is vital for the consystency of the theory. The true check of consistency would be
the absence of negative anticommutators. We have not checked this in a direct way but since we have established that
radiative correction would introduce a term of the form ${\cal L}_{NEK}$, which is known to lead to negative anticommutators,
we can conclude that in the sense of the theorems of Johnson-Sudarshan and Hagen the theory involving interactions
of the form ${\cal L}_P$ is as inconsistent as that involving ${\cal L}_{NEK}$ if we allow electromagnetic interactions.

On the other hand, the very interesting possibility that the interaction ${\cal L}_P$ in absence of electromagnetic
interactions really avoids the above mentioned
consistency problems with the appearance of negative anticommutators, has
not been analyzed with the detail it deserves, and the argumentation given in favor that this inconsistency
is absent has been only sketched. This is interesting at a formal field-theoretic level but irrelevant for
phenomenology, since $\Delta$ resonances are indeed electrically charged. We plan to review this point in the future.

\section{Acknowledgements}

C.B. and A.M. are fellows to CONICET (Argentina) and CCT La Plata, Argentina.

\end{document}